\documentstyle[preprint,aps]{revtex}
\begin{document}
\title{Superluminal Paradox and Neutrino}
\author{Guang-jiong Ni}
\address{Department of Physics, Fudan University, Shanghai,200433, China\\
Department of Physics, Portland State University, Portland, OR 97207 USA}
\date{\today}
\maketitle

\begin{abstract}
Einstein's theory of special relativity(SR) and the principle of causality
imply that the speed of any moving object can not exceed that of light in a
vacuum($c$).However,there were many attempts in literature discussing the
particle moving with speed $u>c$(called as superluminal particle or
tachyon)either in the scheme of SR or beyond it.These theories all
encountered a series of insurmountable difficulties which will be named
``superluminal paradox''in this paper.We will analyze it in some detail and
then prove that the paradox disappears unambiguously in quantum theory,which
is compatible with SR.Most likely,the superluminal particle in real world is
just a kind of known particle,the neutrino.

\end{abstract}

Consider two inertial frames$\sum $ and $\sum^{^{\prime }}$ moving with
relative velocity $v$ along x axis.Then the Lorentz tansformation (LT) reads:

\begin{equation}
x^{^{\prime }}=\frac{x-vt}{\sqrt{1-v^{2}/c^{2}}},t^{^{\prime }}=\frac{%
t-vx/c^{2}}{\sqrt{1-v^{2}/c^{2}}}  \label{1}
\end{equation}%
where ($x,t$) and ($x^{\prime },t^{^{\prime }}$) are the space-time
coordinates of the same particle P in $\sum $ and $\sum^{^{\prime }}$
frames, respectively.

Fig.1 shows the motion trajectory(world line) of a subluminal particle(P)
with uniform velocity $u<c$ on the two-dimensional space-time diagram of $%
\sum $ and $\sum^{^{\prime }}$ .In Fig.1(b) it is seen that when $v>u$,the $%
t^{^{\prime }}$ axis leaps across the world line OP,making $x^{\prime }<0$
and $u^{\prime }=\frac{dx^{\prime }}{dt^{\prime }}<0$ but with $t^{\prime
}>0 $ still. There is no problem[1-4].

However,for a superluminal particle with $u>c$,as shown in Fig.2,a strange
phenomenon occurs.While Fig.2(a) seems not so different from Fig.1.Fig.2(b)
shows a great surprise that when $v>c^2/u$,$t_p^{^{\prime }}$ becomes
negative:

\begin{equation}
t_p^{^{\prime }}<0\text{ }(u>c,v>c^2/u)  \label{2}
\end{equation}
which was regarded as ``tachyon traveling backward in time'' or ``violation
of causality'' and remains as a mysterious puzzle till now.[5-7].

In our opinion,the above puzzle can be better exhibited in an alternative
way.Taking derivative of Eq.(1) ($u=\frac{dx}{dt},u^{\prime }=\frac{%
dx^{\prime }}{dt^{\prime ^{\prime }}}),$we have addition law for velocity in
LT as:

\begin{equation}
u^{\prime }=\frac{u-v}{1-uv/c^{2}}  \label{3}
\end{equation}

Notice that though Eq.(3) is an analytic function of three variables $%
u,u^{^{\prime }}$and $v$ as long as $\left| v\right| <c$ and $u<c$,it does
have a singularity if $u>c$.The pole is located at $v=c^{2}/u$ or $u=c^{2}/v$
and is clearly shown in Fig.3.When $v$ increases across the singularity $%
c^{2}/u$,the velocity of superluminal particle in $\sum^{^{\prime }}$frame,$%
u^{^{\prime }}$,will leap abruptly from $+\infty \rightarrow -\infty :$

\begin{equation}
u^{^{\prime }}<-c,(u>c^2/v\text{ }or\text{ }v>c^2/u)  \label{4}
\end{equation}

Half of above phenomenon can also be seen in Fig.2(a) when we gradually
rotate the x$^{^{\prime }}$axis approaching the OP line anticlockwise:$%
u^{^{\prime }}$=$\frac{\Delta x^{^{\prime }}}{\Delta t^{^{\prime }}}%
\rightarrow \infty .$But it seems to us that Fig.2(b) is meaningless,the $%
x^{^{\prime }}$ axis is not allowed to leap across the OP line.We prefer to
accept Eq.(4) rather than $t_{p}^{^{\prime }}$ 
%TCIMACRO{\TEXTsymbol{<}}%
%BeginExpansion
\mbox{$<$}%
%EndExpansion
0.

However,the EQ.(4) still remains as a puzzle because we have the momentum $%
p^{^{\prime }}>0$ as easily proved from the LT:

\begin{equation}
p^{^{\prime }}=\frac{p-vE/c^2}{\sqrt{1-v^2/c^2}},\text{ }E^{^{\prime }}=%
\frac{E-vp}{\sqrt{1-v^2/c^2}},  \label{5}
\end{equation}
with

\begin{equation}
p=\frac{m_su}{\sqrt{u^2/c^2-1}}>0,\text{ }E=\frac{m_sc^2}{\sqrt{u^2/c^2-1}}%
>0,  \label{6}
\end{equation}
for a superluminal particle(Eq.(6) will be derived
below).Indeed,combining(5) with (6),we find

\begin{equation}
p^{^{\prime }}=\frac{m_s(u-v)}{\sqrt{u^2/c^2-1}\sqrt{1-v^2/c^2}}>m_sc>0.
\label{7}
\end{equation}

\begin{equation}
E^{^{\prime }}=\frac{m_s(c^2-uv)}{\sqrt{u^2/c^2-1}\sqrt{1-v^2/c^2}}%
<0.(u>c^2/v\text{ or }v>c^2/u)  \label{8}
\end{equation}

How can a particle have $u^{^{\prime }}<0$ $(u>c^{2}/v)$ whereas $%
p^{^{\prime }}>0$ ?

Moreover,how can the energy become negative in $\sum^{^{\prime }}$ frame:$%
E^{^{\prime }}<0$ ?(Both $p^{^{\prime }}$ and $E^{^{\prime }}$ vary smoothly
at $uv=c^{2}$ but energy must be positive definite in classical theory.)What
do they mean?All the above puzzle (2),(4),(7) and (8) comprise the
``superluminal paradox''.

Some authors regarded the paradox as a signal showing that the theory of LT
might not be valid for superluminal particles.We don't think so.We believe
the paradox being stemming from the classical nature of above
discussion.Once we find a reasonable quantum theory,the paradox will
disappear.But the clue can only be found from the experiments.

The recent measurements on the neutrino show its mass-square being
negative.Experimental data yield[8](even not so accurately):

\begin{equation}
\frac{E^2}{c^4}-\frac{p^2}{c^2}=m^2(\nu _e)<0  \label{9}
\end{equation}

\begin{equation}
m^2(\nu _e)=-2.5\pm 3.3eV^2  \label{10}
\end{equation}

\begin{equation}
m^2(\nu _u)=-0.016\pm 0.023MeV^2  \label{11}
\end{equation}

Based on these data,one can assume for a superluminal neutrino that:

\begin{equation}
c^2p^2-E^2=m_s^2c^4  \label{12}
\end{equation}
with $m_s^2>0,$for ins$\tan $ce$,m_s(\nu _e)=1.6$ eV.

If accepting Eq.(12),one can easily derive Eq.(6).But how can we derive
Eq.(12) from a quantum theory?In Refs[9,10],a Dirac-type equation is
established,where two two-component spinor functions $\xi (\overrightarrow{x}%
,t)$ and $\eta (\overrightarrow{x},t)$ are coupled together via nonzero $%
m_s: $

\begin{eqnarray}
i\hbar \frac{\partial }{\partial t}\xi &=&ic\hbar \overrightarrow{\sigma }%
\cdot \nabla \xi -m_{s}c^{2}\eta  \label{13} \\
i\hbar \frac{\partial }{\partial t}\eta &=&-ic\hbar \overrightarrow{\sigma }%
\cdot \nabla \eta +m_{s}c^{2}\xi  \nonumber
\end{eqnarray}

Eq.(13) describes the motion of a left-handed neutrino with $E>0$,e.g.,the
plane wave function along x axis being:

\begin{equation}
\xi \sim \eta \sim \exp [\frac i\hbar (px-Et)].\text{ }(\left| \xi /\eta
\right| >1)  \label{14}
\end{equation}

Substitution of (14) into (13) leads to (12) immediately.It is easy to see
that Eq.(13) is invariant under the following ``pure time inversion ''.

\begin{equation}
t\rightarrow -t,\xi (x,-t)\rightarrow \eta _c(x,t),\eta (x,-t)\rightarrow
\xi _c(x,t),  \label{15}
\end{equation}

Meanwhile,the concrete solution (14) is transformed into

\begin{equation}
\eta _c\sim \xi _c\sim \exp [\frac i\hbar (px+Et)].\text{ }(\left| \eta
_c/\xi _c\right| >1)  \label{16}
\end{equation}

If using the familiar momentum and energy operators for particle:

\begin{equation}
\widehat{p}=-i\hbar \frac{\partial }{\partial x},\widehat{E}=i\hbar \frac{%
\partial }{\partial t}  \label{17}
\end{equation}
one would say that Eq.(16) describes a particle with momentum $p$ and energy 
$(-E)<0$.However,a negative-energy particle should be directly viewed as its
antiparticle with positive energy $(E>0)$[11].The counterpart of (17) for
antiparticle read:

\begin{equation}
\widehat{p}_{c}=i\hbar \frac{\partial }{\partial x},\widehat{E}_{c}=-i\hbar 
\frac{\partial }{\partial t}  \label{18}
\end{equation}
with subscript $c$ denoting ``antiparticle''.Hence Eq.(16) should be recast
into

\begin{equation}
\eta _{c}\sim \xi _{c}\sim \exp [-\frac{i}{\hbar }(p_{c}x-E_{c}t)].\text{ }%
(\left| \eta _{c}/\xi _{c}\right| >1)  \label{19}
\end{equation}
with $p_{c}=-p<0,$ and $E_{c}=E>0.$ So Eq.(16) describes a right-handed
antineutrino moving in the opposite direction of x axis.

We are now in a position to solve the ``superluminal paradox'',returning
back to Eqs.(2),(4),(7) and (8).Evidently,the observer in $\sum^{\prime }$
frame will see the neutrino with $u>c^{2}/v$ in $\sum $ frame as an
antineutrino since $E^{\prime }$ $<0$.And its momentum is $p_{c}=-p^{\prime
} $ $<0$(instead of $p^{\prime }>0$) just in comformity with its velocity $%
u^{\prime }<0.$The mysterious time-reversal,Eq(2),is no more than a false
appearance of the sign change in the phase of wave function,which,of course,
can not be reflected suitably in Fig.2(b).So now all puzzle disappear.There
is no paradox at all.

The implication of Fig.3 is amazing.If we tentatively identify the $\sum $
frame with the rest frame of cosmos in which the 3K microwave radiation
background is strictly isotropic,our earth($\sum^{\prime }$ frame)is moving
with velocity $v=356$km/s.Then the originally isotropic neutrinos(identified
tentatively with the dark matter)with velocity distribution in the range $%
(-\infty <u<-c,c<u<\infty )$ will be divided anisotropically in $%
\sum^{\prime }$ frame into two parts.One of them(with $u>c^{2})$ will be
transformed into antineutrinos moving in the opposite direction of $v$ $%
(u^{\prime }<(-c^{2}/v),$see Fig.3(a)$).$As an ideal experiment,if we wish
to chase a superluminal neutrino with fixed $u$ by increasing our velocity $%
v $(see Fig.3(b)),its behavior looks fantastic.First it flees away with
speed $u^{\prime }$ even faster and faster($u^{\prime }>u)$ until $u^{\prime
}=\infty $ when $v\rightarrow c^{2}/u.$Then if we further accelerate to pass
the critical value $v_{c}=c^{2}/u,$it changes suddenly into an antineutrino
moving toward us($u^{\prime }<-c).$(This is why the $x^{\prime }$ can not
leap across the OP line in Fig.2(b)). On the other hand,if we leave the
neutrino along opposite direction ($v<0)$,its velocity ($u^{\prime }>0)$%
slows down instead.

\smallskip

{\bf Summary and discussion:}

(a).Numerous experimental tests have been supporting the validity of
SR,which stands even more firm than ever before.However,based on the new
experimental fact about neutrino,it is possible to construct a superluminal
theory compatible with SR.

(b).In particular,the LT (Eqs.(1) and (5)) and the addition law for velocity
(Eq.(3))are valid for both subluminal and superluminal phenomena as long as $%
\left| v\right| <c$.This is because the concept about space-time is formed
by observers ourselves who are composed of ordinary particles.Our discussion
can be meaningful only if it is based on SR,LT and the invariance of speed
of light in the vacuum($c$).

(c).The superluminal paradox is over.All puzzle stemming from the classical
concepts dissapears in reasonable quantum theory.Indeed,the superluminal
problem poses a very severe and interesting test on the validity of Eq.(13)
which in turn is based on the new concept about the symmetry between
particle and antiparticle(including Eq.(18) vs (17) and Eq.(19) vs (14))
[10,11].

(d).At first sight,the existence of a rest frame $\sum =\sum_{0}$ in cosmos
(implied by the 3K microwave radiation background)leads to vialation of the
symmetry in LT:other inertial frame $\sum^{\prime }$ with velocity $v\neq 0$
relative to $\Sigma _{0}$ is not equivalent to $\sum_{0}$.But since $%
\sum_{0} $ is selected by the neutrino(dark matter)(its velocity
distribution being isotropic in $\sum_{0}$),the equal existence of
antineutrino with mutual transformation between them as shown in Fig.3 does
imply that the LT symmetry is restored implicitly.In other words,The LT
symmetry is hidden in superluminal neutrino bath.The LT symmetry is actually
extended to the totality of inertial frames with velocity in the whole range(%
$-c<v<c$)and $%
%TCIMACRO{\func{real}}%
%BeginExpansion
\mathop{\rm real}%
%EndExpansion
$ized by neutrino and antineutrino together.

The author wishes to thank Dr. T. Chang for bringing the superluminal
problem about neutrino to his attention and relevant discussions.

{\bf Caption:}

Figure 1.A subluminal partcle (P) moving along x axis with velocity $u<c.$

(a)$v<u$; (b)$v>u$. (v is the velocity of $\sum^{\prime }$ frame relative to 
$\sum $.In the limit $v\rightarrow c,x^{\prime }$ and $t^{\prime }$ axes
coincide at the diagonal dash-dot line).

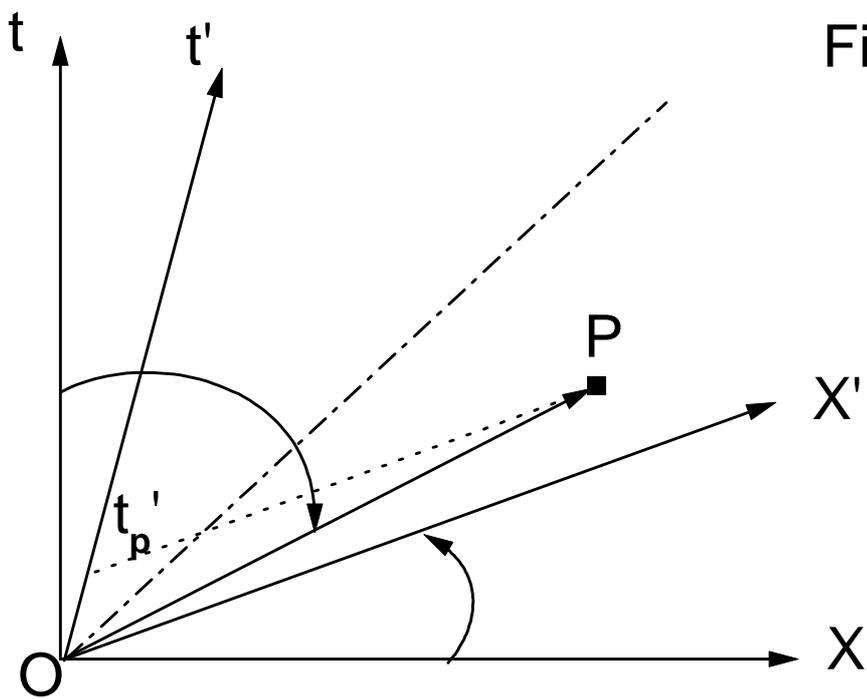
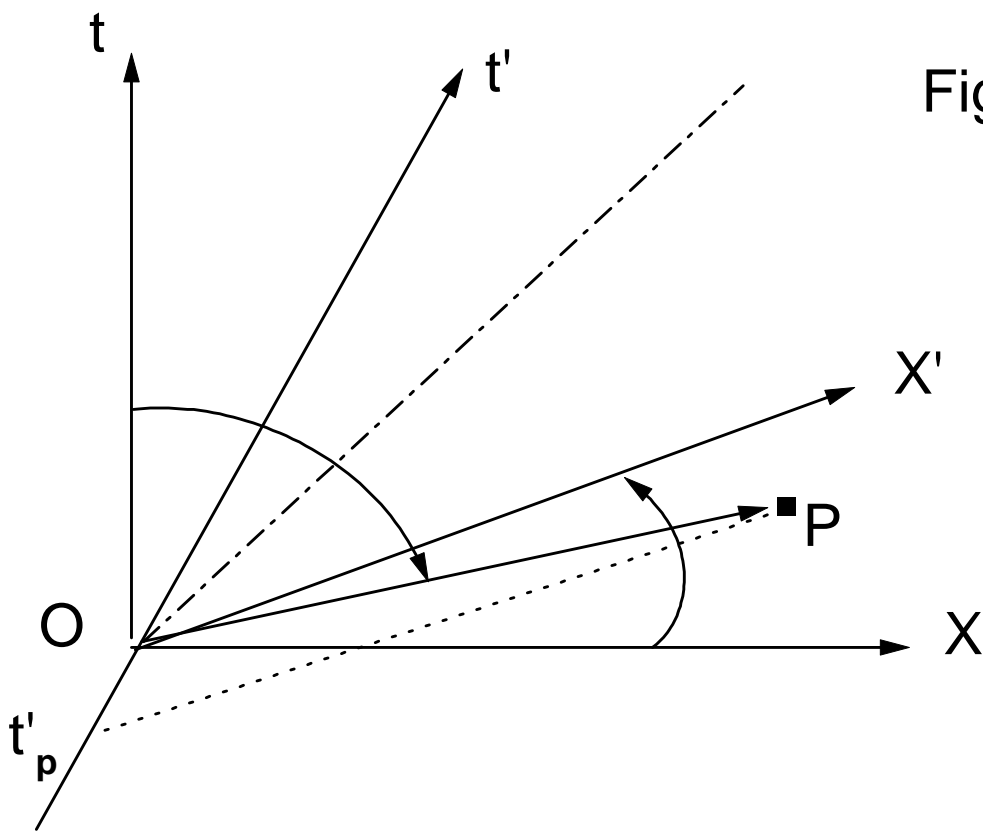
Figure 2.A superluminal particle (P) moving along x axis with velocity $u>c.$

(a)$v<c^2/u,$ $t_p^{\prime }>0;$ $(b)v>c^2/u,$ $t_p^{\prime }<0$

Figure 3.Addition of velocity in Lorentz transformation:

(a) $u^{\prime }$ as a function of $u$ for a fixed $v;$

(b) $u^{\prime }$ as a function of $v$ for a fixed $u(>c)$

\smallskip 

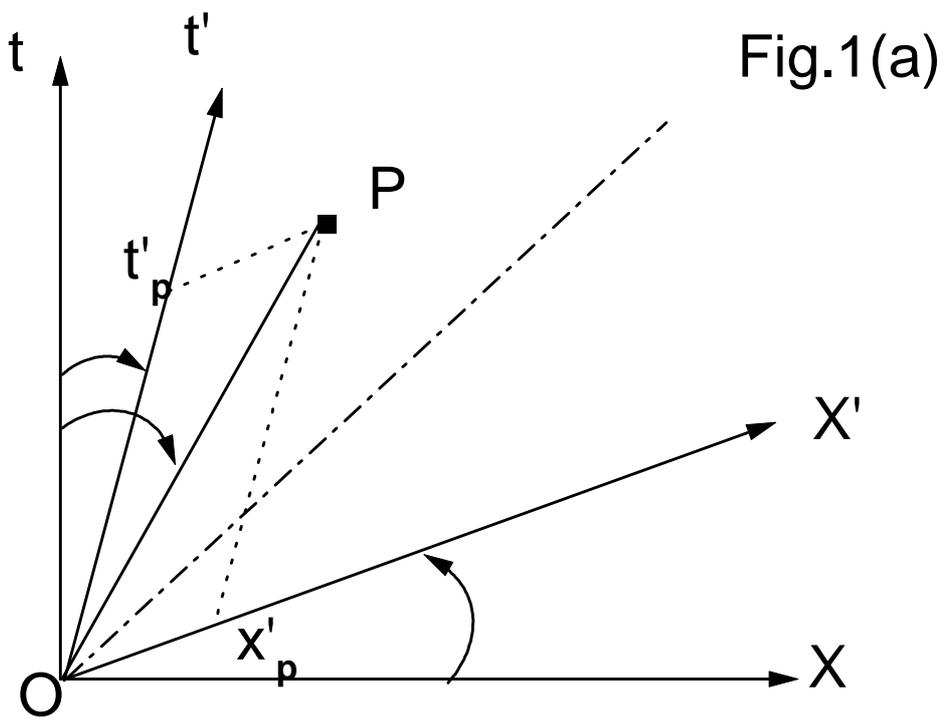
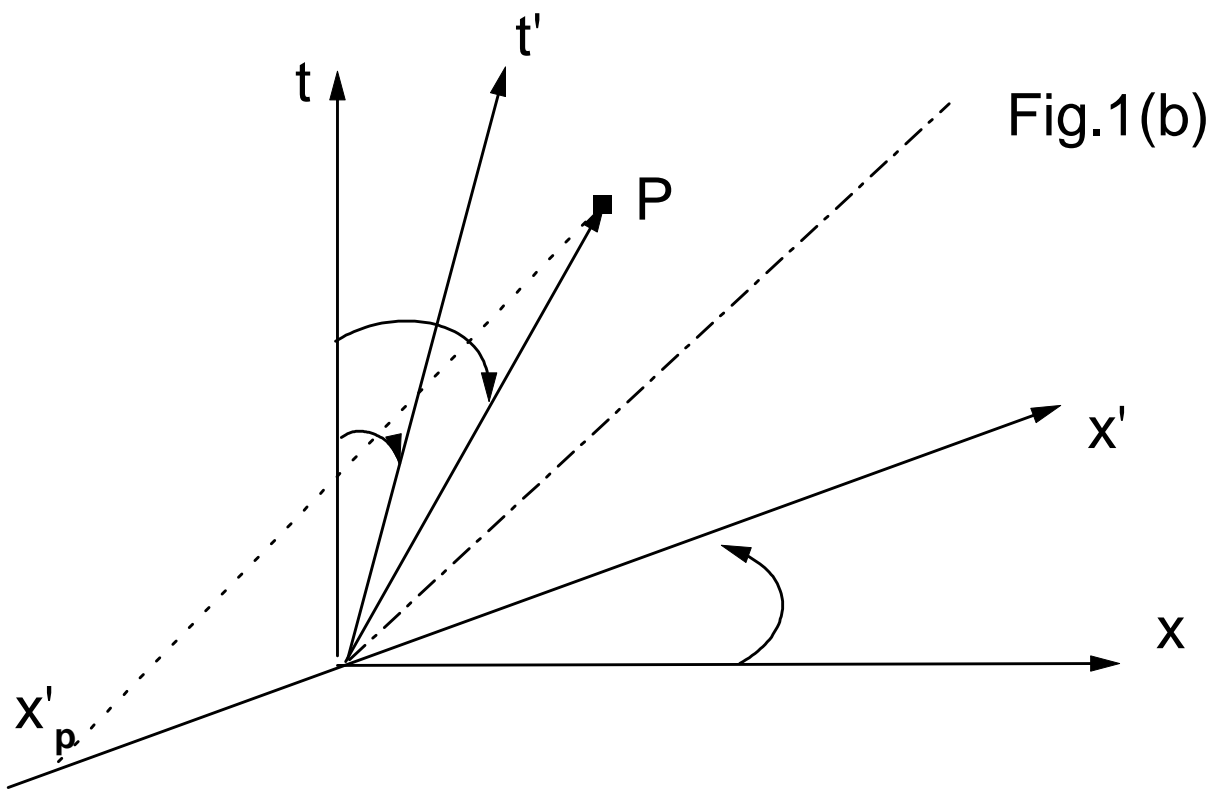
\end{document}